# Single crystal diamond membranes for nanoelectronics

K. Bray[1], H. Kato[2], R. Previdi[1], R. Sandstrom[1], K. Ganesan[3], M. Ogura[2], T. Makino[2], S. Yamasaki[2], A. P. Magyar[4], M. Toth[*1] and I. Aharonovich[1*]

[1.] School of Mathematical and Physical Sciences, University of Technology Sydney, Ultimo, NSW, 2007, Australia

[2] Advanced Power Electronics Research Center, AIST, Tsukuba, Ibaraki, 305-8568, Japan

[3.] School of Physics, University of Melbourne, Vic, 3010, Australia

[4.] Draper Laboratory, Cambridge MA, USA

**Abstract**

**Single crystal, nanoscale diamond membranes are highly sought after for a variety of applications including nanophotonics, nanoelectronics and quantum information science. However, so far, the availability of conductive diamond membranes remained an unreachable goal. In this work we present a complete nanofabrication methodology for engineering high aspect ratio, electrically active single crystal diamond membranes. The membranes have large lateral directions, exceeding ~ 500×500 $\mu m^2$ and are only several hundreds of nanometers thick. We further realize vertical single crystal p-n junctions, made from the diamond membranes that exhibit onset voltages of ~ 10V and a current of several mA. Moreover, we deterministically introduce optically active color centers into the membranes, and demonstrate for the first time a single crystal nanoscale diamond LED. The robust and scalable approach to engineer the electrically active single crystal diamond membranes, offers new pathways for advanced nanophotonics, nanoelectronics and optomechanics devices employing diamond.**

**Introduction**

The majority of modern photonic and optoelectronic devices including photodetectors, light emitting diodes (LEDs), lasers, micro-electromechanical systems (MEMS) and sensors rely on efficient doping (p-type and n-type) for electrical triggering or readout. Many devices also require robust nanofabrication protocols that enable engineering of nanoscale (tens to hundreds of nanometers) suspended membranes. These membranes are fundamental building blocks of advanced nanophotonic components such as waveguides or photonic crystal cavities, and also crucial to enable mechanical motion in MEMs [1-5].

The exceptional properties of diamond, including excellent thermal conductivity, high young modulus and wide optical transparency[6-10], make it an ideal platform for a vast majority of these applications. Diamond is also poised to be the leading candidate for modern nano-electronic devices due to its ability to sustain high temperatures and high electric fields before breakdown[11-13]. However, broad adoption of diamond-based devices has so far been hindered by a lack of large area conductive diamond membranes and efficient nanoscale p-n junctions. While progress has been made to demonstrate proof of concept experiments and develop nanofabrication protocols to sculpt diamond [4, 14-22], large-scale conductive diamond membranes that are suitable for efficient p-n or p-i-n junction engineering are currently beyond reach. This is due, in part, by the challenges associated with epitaxial growth onto non-diamond sacrificial substrates that can be subsequently chemically removed – a process that is well established for silicon and gallium arsenide.

In this work, we overcome these barriers and demonstrate a robust method to fabricate p-type diamond membranes and engineer vertical p-n and p-i-n junctions that are suitable for further development of on-chip nanoelectronic devices. The nanoscale diamond membranes produced in our work have large lateral dimensions, over *500 x 500 μm²*, excellent electronic properties, and are fabricated using a highly robust and technologically mature process, making them attractive candidates for broad adoption and scalable device fabrication.

**Results and Discussion**

To generate the p-type single crystal diamond membranes, we start with boron-doped single crystal diamond grown by chemical vapor deposition (CVD). Boron has been reliably used to achieve p-type doping in diamond[23-24]. The diamond is then implanted with 1 MeV of He$^+$ ions to create a layer of graphitic carbon that can be subsequently removed using an electrochemical etching process that leaves behind standalone diamond membranes with a thickness of 1.7 μm. The process is illustrated in figure 1, and further details are provided in the methods section. Using an optimized electrochemical etch method, we are able to lift off membranes with high surface area. Figure 1f shows an optical image of a single crystal diamond membrane with an area of ~ 0.2 mm² and a thickness of 1.7 μm. Raman spectroscopy was used to characterize the membrane, and a peak at ~1332 cm$^{-1}$ corresponding to a single crystal diamond is clearly seen (Supplementary Information, SI, Figure S1). The membranes were subsequently thinned by Reactive Ion Etching (RIE) to achieve thicknesses smaller than 300 nm. The surface roughness of the membrane before and after etching, was determined by an atomic force microscope to be below 3 nm, suitable for photonic and electronic devices (Figure S1).

To test the electrical properties of the p-type diamond membranes, we measured the current – voltage (I-V) curves for a thick (1.7 μm) and a thinned (~ 200 nm) membrane. An optical image of the latter is shown in figure 2a, while the electrical measurements are presented in figure 2 (b, c). As expected, semiconducting behaviour (Schottky junction) is observed from both membranes. The thicker membrane exhibits a forward threshold voltage of ~10 V and a breakdown voltage of -5V, while the thin membrane exhibits a slightly reduced threshold of 8 V. In both cases, currents on the order of ~ mA (~1.5 – 2 A/cm²) are recorded. These values are comparable to bulk and polycrystalline diamond devices[23, 25]. For reference, the inset of figure 2c shows the same measurement performed on an un-doped diamond membrane with the same thickness (200 nm), that results in a small leakage current of several ~ μA (~2 mA/cm²) flowing through the junction. These results confirm that the boron doped diamond membranes maintain their electrical properties even when they are thinned down to several hundred nanometers.

The availability of p-type membranes enables the exploration of p-n junction engineering using the single crystal diamond membranes. To realize the p-n junctions, we employ a modified fabrication protocol in which an n-type diamond layer is epitaxially overgrown after the helium implantation step. The n-type layer thickness is set to ~ 200 nm, and gas-phase phosphorous is used as the dopant during the CVD growth. Our experimental conditions allow a slow growth that achieves uniform n-type doping across the sample. The overgrown diamond was then processed as described previously to produce a single crystal membrane with a total thickness of ~1.9 μm that is a vertical p-n junction (see figure 3(a-b)). An optical image of the membrane is shown in the inset of figure 3(b). It exhibits diode-like behaviour with a forward threshold voltage of ~ 10 V with ~ mA (~ 3A/cm²) current, and a negligible current (~1×10$^{-7}$ A/cm²) under a reverse bias of > 40V (figure 3c). This leads to an extremely high rectification ratio of ~ 10$^4$ (semi log I-V plots are shown in SI) which is ideal for diode fabrication[26-27]. This value is comparable to bulk diamond devces[23].

Upon subsequent thinning of the membrane to ~ 300 nm by RIE (yielding a p-type layer of ~ 100 nm and an n-type layer of ~200 nm), p-n membrane still behaves as a diode with a forward voltage of ~ 8 V with ~ mA (~ 2 A/cm$^2$) current, and negligible current (~2x10$^{-7}$ A/cm$^2$) under reverse bias (figure 3d). The rectification ratio of the thin membrane is worse, ~ 30, and may be caused by roughening during the RIE process or by heterogeneous boron doping near the p-n interface. To the best of our knowledge, this is the first report of a vertical nanoscale p-n junction engineered within single crystal diamond, and is highly promising for a myriad of nano-electronic applications.

To determine the thickness of the depletion region, $\omega$, of the p-n diamond junction, the following equation was employed,

$$\omega = \sqrt{\frac{2\varepsilon_s}{q}(\frac{1}{N_A} + \frac{1}{N_D})(\phi_i - V_a)} \qquad \text{Equation 1.}$$

where $\varepsilon_s$ is the permittivity, $q$ is the electron charge, $N_A$ and $N_B$ are the acceptor ([B]~3×10$^{20}$/cm$^3$) and donor ([P]~8x10$^{18}$/cm$^3$) concentrations, respectively, $\phi_i$ is the built-in potential (~ 5.5 V) of our device and $V_a$ is the applied voltage (calculation is done for $V_a = 0$ at equilibrium). Under our experimental conditions, a depletion layer of ~ 21 nm exists. The short depletion layer indicates that only a minimal voltage is needed to overcome this barrier potential and allow for recombination of electrons and holes flowing from the n$^-$ and p$^+$ layers respectively and is important for practical devices. Furthermore, the p-n single crystal diamond membrane devices show a maximum electric field ($\mathcal{E} = 2(\phi_a - V_i)/w$) of ~ 500 MV/m. While this field is not as high as millimetre scale bulk diamond, it is still an order of magnitude greater than silicon (300 kV/m) which is ordinarily used as the base material for nanoelectronic devices. A higher electric field allows the diamond membranes to withstand higher voltages (up to ~ 100 V for a 500 nm vertical device) and stresses before the device electrically breaks down and the material becomes conductive. Overall, the short depletion layer and the high breakdown field makes these nanoscale single crystal diamond membranes extremely attractive for real nanoelectronic devices.

Beyond nanoelectronics, we foresee the great potential of our diamond membranes for optoelectronic and nanophotonic applications. To demonstrate a working device, we fabricated a p-i-n structure into a single crystal diamond membrane containing silicon vacancy (SiV) color centers that display both photoluminescence (PL) and electroluminescence (EL) (figure 4). The inclusion of the SiVs into the intrinsic layer between the p- and the n-doped regions allows for the recombination process to occur at the SiV centers, thus generating EL signal. The SiVs color centers were chosen intentionally, as they are emerging fluorescent defects in diamond that hold great potential for myriad of quantum photonic applications.[3, 28] To realize this structure, a boron doped implanted single crystal diamond was overgrown in the presence of a silicon source to introduce the SiV color centers (into the intrinsic layer). An n-type diamond was subsequently epitaxially overgrown and the membrane was lifted off as described previously.

An optical image of the stand-alone p-i-n single crystal membrane device is shown in figure 4a and the schematic illustration of the structure and the device are shown in figure 4b. A schematic energy band diagram is shown in the SI. The formed single crystal diamond membrane device displays excellent electrical characteristics, as shown in figure 4c. Diode-like behaviour with a threshold voltage at ~11 V and negligible current under a reverse bias of > 20V is measured. This is confirmed by the semi-log I-V curve with a high rectification ratio of ~ 10$^6$ (inset of figure 4c). The low threshold voltage is advantageous for efficient device operation and is comparable with many standard nano-optoelectronic devices made of GaN, SiC and ZnO[29-31].

Figure 4d shows the PL and the EL characteristics of the device, recorded at room temperature. The PL was recorded using a 532 nm laser excitation, while the EL was recorded under a forward bias with an injection current of 3 mA. Both the EL and the PL exhibit the desirable emission of the SiV color centers at ~ 737 nm.

We now discuss our results in the context of diamond membranes performance and compare it to the results reported by other research groups. Table 1 summarizes the literature of diamond membranes and provides the key characteristics of size and optical activity. Notably, the membranes reported in this work are amongst the largest in terms of surface area, while still being relatively thin – on the order of ~ 300 nm thick. Furthermore, our membranes have the advantage of being free standing single crystal diamond that can be transferred and accurately positioned on a substrate of choice. The membranes engineered in the current work also host optically active defects that can be triggered both optically and electrically. All these attributes combined are unique, and ideal for future fabrication of integrated nanophotonic, optoelectronic and optomechanical circuits using single crystal diamond membranes.

**Conclusions**

To conclude, we show the first reliable and robust engineering of nanoscale conductive diamond membranes. Moreover, we engineer the first-of-its-kind vertical nanoscale p-n and p-i-n nanoscale devices using entirely single crystal diamond. The membranes and the devices have high aspect ratio of lateral to vertical dimensions, are standalone and can be positioned onto a substrate of choice. Finally, we show that the p-i-n single crystal diamond devices can host optically active color centers – namely the SiV centers, that exhibit excellent EL and PL characteristics, and therefore are suitable for optoelectronic and photonic applications.

Our work paves the way for numerous exciting new avenues employing the electrically active membranes. For instance, scalable integrated quantum nanophotonics circuits – including waveguides and photonic crystal cavities with electrically driven single photon emitters may become possible[32-33]. The ability of diamond to host myriad of color centers, can be leveraged to realize arrays of nanoscale multicolour LEDs that operate under harsh chemical and physical environments – as diamond is chemically inert. In addition, advanced nano-electromechanical systems (NEMS)[34] may be realized. The availability of robust single crystal p-n junctions may lead to engineering of nanoscale diamond field effect transistors with high breakdown voltages to achieve fast nanoelectronic circuits on a single chip. Finally, advanced sensing techniques – particularly for DNA translocation[35] where excellent mechanical and electrical properties are needed – may be developed.

**Methods**

**Fabrication of p-type diamond membranes.**

Boron doped membranes were created from a bulk boron doped (~ $3 \times 10^{20}$ atoms/cm$^3$) diamond crystal grown by Microwave Plasma Chemical Vapor Deposition (MPCVD). The crystal was implanted with 1 MeV He$^+$ ions to a dose of $5 \times 10^{16}$ ions/cm$^2$ and subsequently annealed at 900 $^\text{o}$C in vacuum to create a thin amorphous carbon layer 1.7 μm below the surface, as shown in figure 1. The amorphous layer enables lift-off of the 1.7 μm thick diamond membrane by electrochemical etching. To selectively remove the graphitic residual, the sample was then immersed in deionized water and electrochemical etching was carried out using a constant forward bias of 60 V. During etching, a positively biased tip was contacted to the top surface of the diamond while the negatively biased tip was positioned slightly above the substrate. The diamond membranes were cleaned using a 3:1 (sulphuric acid - hydrogen peroxide) piranha solution and transferred using a liquid droplet to a silicon substrate coated with ~150 nm of titanium which acts as a sticking layer and a bottom contact. The membranes range in size from [~50 to 1000] μm × [~100 to 2000] μm.

A DektaXT stylus Profilometer was used to determine the thickness of the diamond membranes. To thin the diamond membranes, an inductive coupled plasma reactive ion etching (RIE) with a tetrafluoromethane/oxygen ($CF_4/O_2$) ratio of 1:3 at a pressure of 20 Pa, with a forward power of 200 W was used.

**Fabrication of vertical nano-scale p-n junctions.**
A 200 nm n-layer was grown epitaxially using MPCVD on top of $He^+$ ion implanted boron doped single crystal bulk diamond, before the electrochemical etch. For the n-type layer, $PH_3$ diluted with $H_2$ was used for phosphorus doping at a ratio ($PH_3/CH_4$) of 5% and the final doping is $\sim 8 \times 10^{18}$ cm$^{-3}$ of phosphorous atoms. The sample then underwent electrochemical etching to dissociate the p-n diamond membrane that is ~1.9 μm thick. The p-layer was thinned by RIE to achieve a total membrane thickness of ~300 nm.

**Fabrication of vertical p-i-n junction.**
To fabricate the vertical p-i-n junction, a 100 nm intrinsic layer was grown epitaxially on top of $He^+$ ion implanted boron doped single crystal bulk diamond, following by a 200 nm n-layer [~ $8\times10^{18}$ cm$^{-3}$ phosphorous]. The sample then underwent electrochemical etching to dissociate the p-i-n diamond membrane.

**Fabrication of vertical p-i-ITO junction.**
A ~200 nm boron doped p-type diamond membrane was fabricated as explained above and positioned onto a clean silicon substrate. The membrane was then overgrown using a MPCVD chamber with a hydrogen/methane ratio of 100:1 at 60 Torr, a microwave power of 900W for 8 minutes to fabricate an ~100 nm intrinsic diamond layer that contains SiV color centers. The silicon doping occurs naturally with the silicon source from the substrate incorporates into the membrane. The p-i membrane was subsequently transferred onto a titanium coated silicon and sputtered with ~100 nm of n-doped indium titanium oxide (ITO) through a photolithographic mask.

**Electrical measurements.**
For the electrical measurements, the membranes were positioned onto a titanium (~150 nm) coated silicon dioxide wafer. The top contact was titanium/gold metal that was sputtered through a lithographic mask (thickness ~ 150 nm and 20 μm diameter). To contact the devices, two xyz nanomanipulators were used, one positioned on top of the membrane and the second one positioned in a close proximity to it. The voltage was generated and controlled using the Keithley 617 electrometer and the current passing through the device was measured using the same tool. The data was recorded using a custom built software. All the measurements were done at room temperature.

**Optical Characterization.**

A continuous-wave 532 nm laser (Gem 532, Laser Quantum) was used for excitation, focused onto the sample using a high numerical-aperture (NA = 0.9, Nikon) objective lens. The collected light was filtered using a 532 nm dichroic mirror (532 nm laser BrightLine, Semrock) and an additional long-pass filter (Semrock). The signal was then coupled into a graded-index fibre, with the fibre aperture serving as a confocal pinhole, onto a spectrometer (Princeton Instruments).

**Acknowledgements**

We would like to acknowledge Carlo Bradac, Amanuel Berhane and Blake Regan for useful discussions, Minh Anh Phan Nguyen for AFM measurements and John Scott for TEM measurements and analysis. Financial support from the Australian Research Council (via DP140102721, DE130100592), FEI Company, the Asian Office of Aerospace Research and Development grant FA2386-15-1-4044 are gratefully acknowledged. I.A. gratefully acknowledges JSPS Invitation Fellowships [S16712]. This research is supported by an Australian Government Research Training Program Scholarship.

**Competing interests:** The authors declare that they have no competing interests.

**Figures**

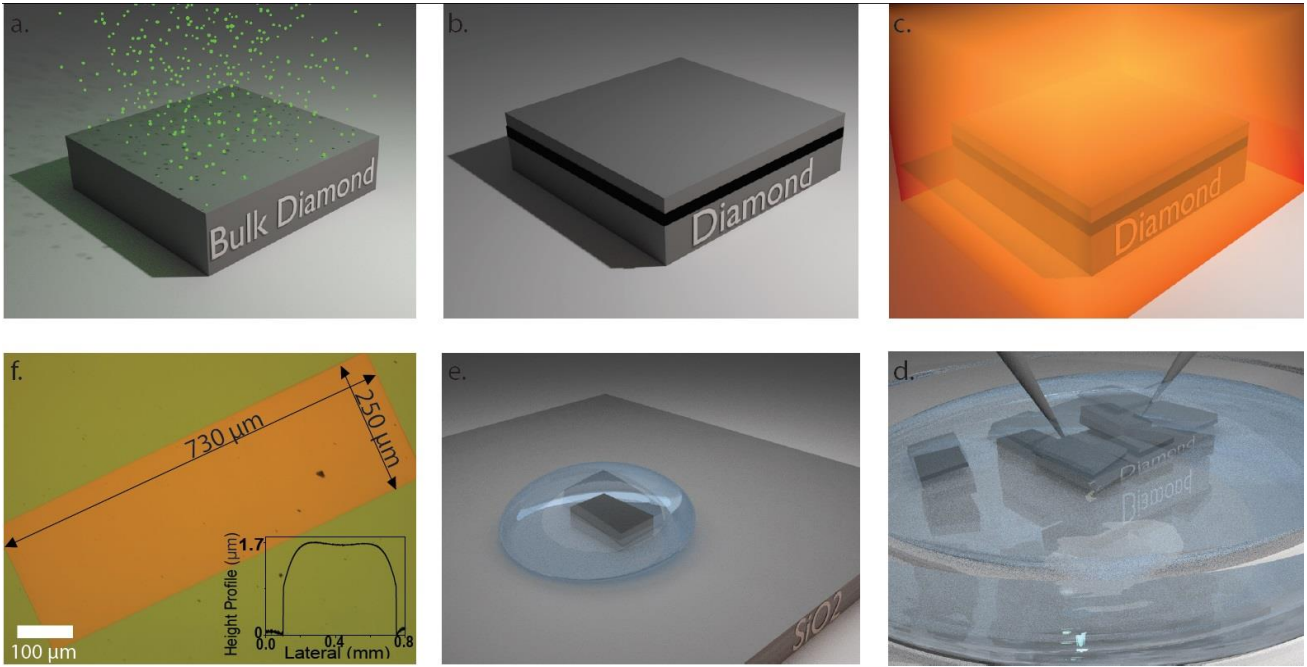

**Figure 1.** Schematic of the process used to engineer single crystal conductive diamond membranes. (a) Single crystal, boron doped CVD diamond is implanted with He ions (1 MeV, $5 \times 10^{16}$ ions/cm$^2$) to create a damaged layer ~ 1.7 µm below the surface (b). (c) The sample is annealed at 900$^0$ C and (d) electrochemically etched to lift off the diamond membranes. (e) The membranes are transferred onto a Ti coated substrate using a liquid transfer process. (f) A top-down optical image (false color) of a membrane that has been lifted off. Inset: height profile that shows the membrane is ~ 1.7 µm thick.

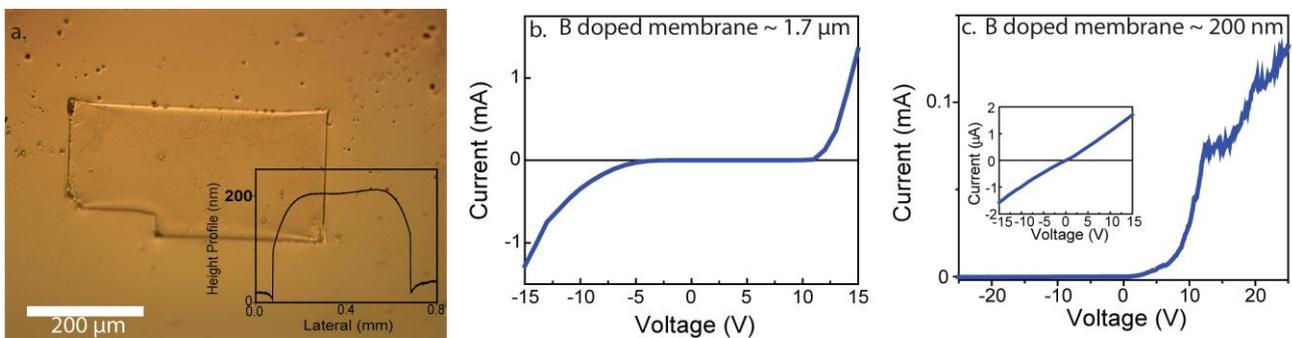

**Figure 2.** Boron-doped diamond membranes. (a) Optical image of a thin boron doped diamond membrane. Inset: height profile that confirms the membrane is ~ 200 nm thick. (b) An I-V curve of the lifted off boron doped membrane (~1.7 µm) exhibits a threshold of ~ 10 V. (c) An I-V curve of a 200 nm thick, boron doped membrane exhibits an on-set threshold of ~ 8 V. Both I-V curves show semiconducting behaviour. Inset: IV curve of an undoped diamond membrane showing insulating behaviour with negligible µA currents.

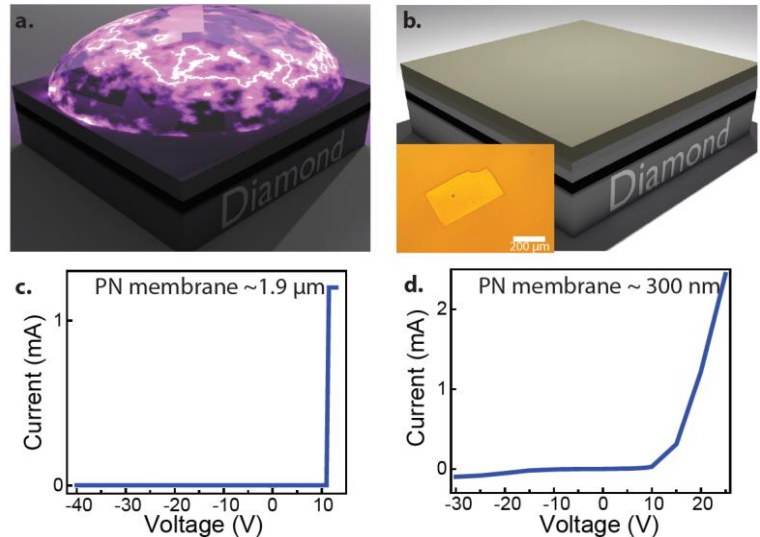

**Figure 3.** Engineering of vertical p-n junctions. (a-b) Schematic of the process used to produce single crystal p-n diamond membranes. (a, b) Bulk boron doped diamond that was implanted with He ions to create a graphitic channel is overgrown with diamond to create a phosphorus doped n-type layer represented by the dark green color in (b). Inset, the lifted-off entire p-n diamond membranes. (c) Electrical measurements of the lifted off, 1.9 μm thick PN membrane showing diode behaviour at ~10 V and (f) an I-V curve of the thinned (~300 nm) p-n membrane, showing a similar on-set voltage.

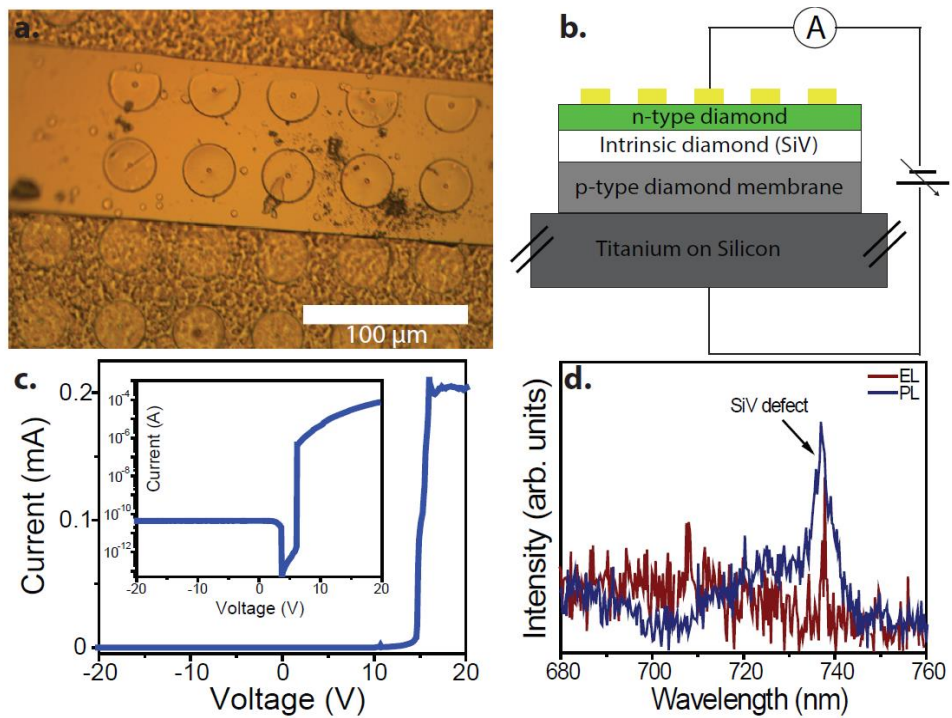

**Figure 4.** Vertical standalone single crystal diamond p-i-n device. (a) An optical image of the p-i-n membrane containing a layer of SiV emitters in the intrinsic region. The circles are the metal contacts on the top n-type surface. (b) Schematic diagram of the cross section of the p-i-n diamond device. (c) The p-i-n membrane shows diode-like behaviour with an on-set voltage of ~ 11V. Inset: Log (I) vs (V) rectification curve of the p-i-n diamond membrane showing a rectification ratio of ~ $10^6$. (d) Photoluminescence (blue) and electroluminescence (red) measurements recorded at room temperature showing a characteristic emission from the same SiV color center at ~737 nm.

| Length×Width (μm) | Thickness (nm) | PL | EL | Free standing | Ref |
|---|---|---|---|---|---|
| 15 x 3 | 200 | No | No | No | 36 |
| 2000 x 2000 | 3000 | No | No | No | 37 |
| 3000 x 3000 | 2000 | No | No | No | 38 |
| 400 x 400 | 1000 | Yes, (NV) | No | No* | 15, 39 |
| 1100 (diameter) | 1200 | Yes, (NV) | No | No | 40 |
| 300 x 300 | 200 | No | No | Yes | 41 |
| 4500 x 4500 | 15000 | No | No | Yes | 42 |
| 10 x 10 | 200 | Yes, (NV) | No | Yes | 43 |
| 220 x 220 | 200 | Yes (NV, SiV) | No | Yes | 17, 44 |
| 1000 x 500 | 300 | Yes, (SiV) | Yes, (SiV) | Yes | Our work |

**Table 1.** A table *summarising the performance* of diamond membranes. *The main parameters of size, optical activity (photoluminescence), electroluminescence (EL) and whether the membrane is free standing or attached are provided.* *The membrane was attached to a large polycrystalline diamond frame